\documentclass[conference]{IEEEtran}
\usepackage{cite}
\usepackage{amsmath}
\usepackage{graphicx}
\graphicspath{{figures/}}
\usepackage{booktabs}
\usepackage{multirow}
\usepackage{threeparttable}
\usepackage[table]{xcolor}
\usepackage{listings}
\usepackage{tcolorbox}

\providecommand{\rqanswer}[2]{%
  \begin{tcolorbox}[colback=black!4, colframe=black!55, boxrule=0.5pt, arc=1mm,
    left=5pt, right=5pt, top=3pt, bottom=3pt, fontupper=\small]
  \textbf{Answer to #1.} #2
  \end{tcolorbox}}
\usepackage[hidelinks]{hyperref}

\begin{document}

\title{Safe by Accident: Crash-Testing LLM-Written Code}


\author{
\IEEEauthorblockN{Afridi Mahmud}
\IEEEauthorblockA{
Bangladesh University of\\
Engineering and Technology\\
Dhaka, Bangladesh\\
\texttt{afridimahmud99@gmail.com}
}
\and
\IEEEauthorblockN{Md. As-Aid Rahman}
\IEEEauthorblockA{
Bangladesh University of\\
Engineering and Technology\\
Dhaka, Bangladesh\\
\texttt{mdasaidrahman@gmail.com}
}
\and
\IEEEauthorblockN{Rifat Shahriyar}
\IEEEauthorblockA{
Bangladesh University of\\
Engineering and Technology\\
Dhaka, Bangladesh\\
\texttt{rifat@cse.buet.ac.bd}
}
}

\maketitle

\begin{abstract}
Evaluations of LLM-written code run it once, in one environment, and credit the code with
whatever passes. We show that this credit can belong to the environment. We crashed LLM-written
backend code just before each database, file and email call, re-ran it as an operator would, and
judged the result under two transaction modes of Python's \texttt{sqlite3}: the default, in which
the library quietly opens a transaction, and autocommit, in which it does not. Across 9,100
solutions from ten models on 24 backend tasks, more than half (54\%) of the database code that
survives every crash in the default mode is left half-updated in autocommit mode. This holds for
every model, and it hides in ordinary code, including \texttt{with conn:} blocks that look like
transactions but open none. Telling the model that its connection is in autocommit mode does not
change what the local code models write, while the API models' unsafe share falls by 30 points.
Asking for crash safety helps every model, and strong models repair nearly all unsafe code, but
often write safe code only when asked. Thus, much of the crash safety that LLM-written database
code appears to have is borrowed from the library rather than designed into it, and a second test
under a less forgiving configuration reveals the difference.
\end{abstract}
\section{Introduction}

\begin{figure}[t]
\centering
\includegraphics[width=\linewidth]{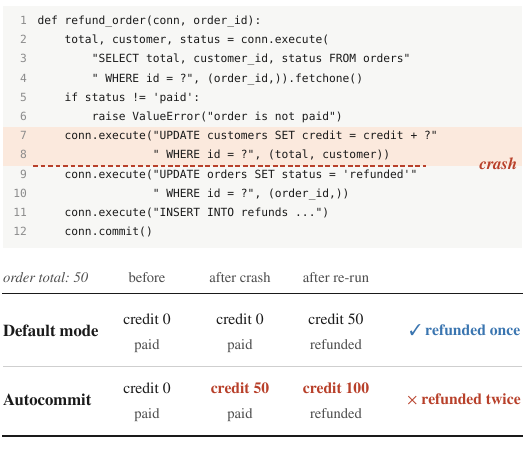}
\caption{An LLM-written refund, crashed after lines 7--8 and re-run. In Python's default mode the customer is refunded once; in autocommit mode, twice.}
\label{fig:intro}
\end{figure}

Consider a small function from an online shop. A customer asks for a refund, and an LLM is
asked to write the code. Figure~\ref{fig:intro} shows what it produces.

The code is short, readable and correct. It passes every functional test. Now suppose the
server is restarted at the point marked in the figure, just after lines 7--8, when the customer
has received the credit but the order has not yet been marked as refunded. An operator notices
the failed job and runs it again. What happens next depends on something the code never
mentions.

In Python's default mode, the database library silently opens a transaction before the first
write and keeps it open until \texttt{commit()}. The crash discards the half-finished refund,
the re-run starts from a clean slate, and the customer is refunded exactly once. But many
applications open their connections in \emph{autocommit} mode, where every statement is saved
the moment it runs. There, the crash leaves the credit in place and the order still marked
``paid'', and the re-run refunds the customer a second time.

A crash test run in the default mode would pass this code. The pass is real, but it belongs to
the library, not to the code. The code is \emph{safe by accident}.

This matters because LLM code is judged by running it. Evaluations check whether it works and,
increasingly, whether it is secure~\cite{pearce2022asleep,vero2025baxbench}, always by running it
once, in one fixed environment. Such a verdict cannot say whether what it observed belongs to the
code or to the environment. When a library default quietly supplies a guarantee, the code is
credited with it, and loses it as soon as a caller sets things up differently.

Crash safety is a sharp place to study this. Jobs are killed by deployments, timeouts and
out-of-memory errors, and payment and email workers are retried as a matter of routine. Code that
leaves half-written data or repeats its work after a crash causes bugs that are rare, expensive
and hard to trace: a double charge, a duplicate email, an account that is neither open nor
closed. The guarantee that prevents this, a transaction around a group of writes, is exactly the
kind a library can supply without the code asking. Its absence is also hard to spot by reading: a
\texttt{with conn:} block, which most reviewers would read as a transaction, opens none in
autocommit mode.

Crash-consistency research crashes software on purpose, but it studies mature, human-written
systems~\cite{pillai2014alice,mohan2018crashmonkey}, and recent work on LLM agents injects faults
into their live tool calls~\cite{li2026exactlyonce}. To our knowledge, no study has crashed the
application code an LLM writes and judged what an operator finds after re-running it, and none
has asked whether a passing verdict belongs to the code or to the environment it ran in. Our key
idea is simple: run the same crash test under two configurations that differ only in a guarantee
the library supplies. Code that passes both earned its pass; code that passes only one borrowed
it. Because both verdicts concern the same solution, the comparison needs neither a human
baseline nor matched models.

We built a harness that crashes each LLM-written solution just before each of its writes,
re-runs it as an operator would, and does both in two transaction modes, and used it on 9,100
solutions from ten models on 24 everyday backend tasks. Our question concerns evaluation, not
authorship: we ask what a passing test credits, not whether models write worse code than people,
whose code may lean on the same default.

We investigate three research questions:
\begin{itemize}
    \item \textbf{RQ1: Borrowed safety.} How much of the apparent crash safety of LLM-written
    code is supplied by the environment rather than by the code?
    \item \textbf{RQ2: Information.} Does telling the model about its environment remove the
    dependence?
    \item \textbf{RQ3: Making safety the code's own.} Does asking for crash safety, or repairing
    unsafe code, produce code that is safe by design?
\end{itemize}

This paper makes four contributions:
\begin{itemize}
    \item \textbf{Safety by accident, and a test that exposes it.} With this two-configuration
    test, we show that more than half of the database code that survives crashes, in every
    model, is safe only because of the library default (RQ1).
    \item \textbf{Where it hides.} In code with no transaction control, and in \texttt{with
    conn:} blocks that look transactional but are not (RQ1).
    \item \textbf{What removes it.} Stating the mode does not change the local code models'
    code; asking for crash safety helps every model, and strong models repair almost all unsafe
    code (RQ2, RQ3).
    \item \textbf{A benchmark and harness.} We release 24 tasks, a deterministic crash harness
    validated against 58 reference solutions, and all 9,100 generated solutions and
    results.
\end{itemize}

\section{Background}
\label{sec:background}

\textbf{Crash safety.} A function is \emph{crash-safe} if killing it at any point, and then
running it again, leaves the data correct. ``Correct'' means two things. The data must never be
left \emph{half-updated}, such as money taken from one account but not added to the other. And
the re-run must not repeat work that already happened, such as a second refund or a second
email; we call this \emph{double-applying}. A transaction prevents the first by saving all of a
group of writes or none; recording what has been done prevents the second by letting a re-run
skip it. Work that leaves the database, like an email, needs the second: no transaction can take
back an email that has been sent.

\textbf{Two transaction modes.} In its default mode, Python's \texttt{sqlite3} module opens a
transaction before the first \texttt{INSERT}, \texttt{UPDATE} or \texttt{DELETE} and keeps it
open until \texttt{commit()}~\cite{python-sqlite3}. In autocommit mode
(\texttt{isolation\_level=None}), each statement is saved at once unless the code opens a
transaction itself. Both modes are mainstream. Django runs its connections in autocommit mode by
default~\cite{django-transactions}, on SQLite by setting exactly
\texttt{isolation\_level=None}~\cite{django-sqlite-backend}; psycopg, the common PostgreSQL
driver, opens a transaction implicitly, as \texttt{sqlite3} does~\cite{psycopg-transactions}; and
Python's documentation announces a future default that keeps a transaction open at all
times~\cite{python-sqlite3}. Which mode a test uses is thus a choice, not a fixed fact about
deployment. We use Python~3.13 and SQLite~3.45.

\textbf{Classes.} A solution \emph{works} if its normal run gives the right result in the default
mode. We classify working code by whether its writes are all-or-nothing, the property a
transaction protects, judged on the data right after each crash:
\begin{itemize}
\item \textbf{Unsafe:} some crash leaves the data half-updated in the default mode.
\item \textbf{Safe by accident:} never half-updated in the default mode, but half-updated in
autocommit mode, or not working there at all.
\item \textbf{Safe by design:} never half-updated in either mode.
\end{itemize}
Repeated work is counted separately, so code that is safe by design can still double-apply. We
also report the share that is \emph{fully crash-safe}: it works in both modes, and after any
crash in either mode a re-run finishes the work exactly once.

Eight tasks are designed to finish interrupted work, such as a resumable import or an email that
may be sent late (marked in Table~\ref{tab:tasks}). For these the re-run is part of the job, so
a re-run that fails, or that repeats work such as sending a queued email twice, counts as unsafe.
For the other tasks the class rests on the data right after the crash, and a state in which
nothing is broken, such as an orphaned file that no record points to, is not half-updated. File
tasks have no transaction default, so they are either unsafe or safe by design.

\textbf{What ``safe by accident'' means.} Each task is a job an operator re-runs, so the function is the whole unit of work, yet
it gets \texttt{conn} from its caller and cannot know its mode. Code that calls \texttt{commit()}
already owns the transaction's end, so it must also open it; code that does so is never safe by
accident (Section~\ref{sec:rq1}). Our references always open it; opening it only when none is
open, which composes inside a caller's transaction, counts the same. The plain prompt names
neither crashes, the mode nor who owns the transaction, so this is no fault of the model:
reasonable code passes a default-mode test for a reason outside itself. Under the prompts that
state the requirement or the mode (Section~\ref{sec:prompts}), an unsafe verdict is a request not
met.

\section{Methodology}
\label{sec:method}

\subsection{Tasks}
We wrote 24 tasks, each a short request of the kind a developer would give a coding assistant
(Table~\ref{tab:tasks}). Nineteen use a SQLite database, eight of them together with an outside
effect such as an email or a file, and five use only files. Each prompt gives the function
signature, the schema and the business rule, and nothing about crashes. Where a real API would
offer protection, the prompt says so: the email client accepts an idempotency key and drops
repeated sends, so code may send before recording if it passes a stable key. Each solution comes
from a single prompt, with no tests, tools or second attempt. The task set was frozen before the
main study, and no task was dropped. The tasks cover common backend work rather than a sample of
real code, so absolute rates describe this task mix.

Every task can be solved safely. Each has a reference solution that the harness classifies as
safe by design, and a repeat is always recognisable, through a request identifier (six tasks) or
the stored state (the rest). The one exception, counting a page view, is not judged on double
counting. Before every campaign the harness must classify all 58 reference solutions as intended:
24 safe by design, 11 safe by accident, 22 unsafe, and 1 that undoes its work when an error is
raised but not when the process is killed, so it must be unsafe only under a hard crash.

\begin{table}[t]
\centering
\begin{threeparttable}
\caption{The 24 tasks. Each is an everyday backend request; none mentions crashes.}
\label{tab:tasks}
\small
\setlength{\tabcolsep}{4pt}
\renewcommand{\arraystretch}{1.12}
\newcommand{\rc}{$^\dagger$}
\newcommand{\nm}[1]{\,{\scriptsize\textcolor{black!55}{#1}}}
\newcommand{\fx}[1]{\textcolor{black!55}{\footnotesize\,+\,#1}}
\begin{tabular*}{\columnwidth}{@{\extracolsep{\fill}}ll@{}}
\toprule
\rowcolor{black!8}
\multicolumn{2}{@{}l}{\textbf{Database only} (11)} \\
Add a comment \nm{12/76}               & Process pending jobs\rc \nm{2/95} \\
Add a slug column\rc \nm{53/0}         & Receive a stock delivery \nm{4/82} \\
Archive old orders \nm{1/63}           & Redeem a gift card \nm{1/55} \\
Delete a user's data \nm{0/77}         & Refund an order \nm{0/74} \\
Import customers (CSV)\rc \nm{9/91}    & Transfer money \nm{0/39} \\
Place an order \nm{1/53}               & \\
\addlinespace[3pt]
\rowcolor{black!8}
\multicolumn{2}{@{}l}{\textbf{Database and an outside effect} (8)} \\
Cancel a subscription\rc \fx{email} \nm{36/3} & Mark an order paid\rc \fx{email} \nm{42/0} \\
Delete an attachment \fx{file} \nm{48/0}      & Send queued emails\rc \fx{email} \nm{24/0} \\
Export a sales report \fx{file} \nm{89/0}     & Store an upload\rc \fx{file} \nm{100/0} \\
Invite a user\rc \fx{email} \nm{64/0}         & Update prices \fx{cache} \nm{75/7} \\
\addlinespace[3pt]
\rowcolor{black!8}
\multicolumn{2}{@{}l}{\textbf{Files only} (5)} \\
Append to a ledger \nm{1}              & Save a user setting \nm{79} \\
Archive imported files \nm{80}         & Set a config option \nm{91} \\
Count a page view \nm{90}              & \\
\bottomrule
\end{tabular*}
\begin{tablenotes}[flushleft]
\footnotesize
\item ``+'': effect outside the database, which no transaction can undo. Small numbers: \% of
working plain-prompt code that is unsafe / safe by accident.
\item $^\dagger$\,Finishes interrupted work; judged after the re-run.
\end{tablenotes}
\end{threeparttable}
\end{table}

\subsection{Models and prompts}
\label{sec:prompts}
We tested ten models (Table~\ref{tab:models}): six \emph{local code models} that we ran on our
own GPUs, and four \emph{API models}. The two groups differ in several ways at once: the API
models are general-purpose, mostly larger and newer, and reason by default, while three local
models are 4-bit quantized. We therefore read group differences descriptively and test single
factors directly: reasoning with Gemma~4 run with thinking switched off, quantization with
Qwen2.5-Coder-7B at both precisions, and size with a near-matched pair, Qwen3-Coder-30B-A3B and
Gemma~4 26B-A4B (3B and 4B active parameters). Reasoning runs at each model's default, except
Flash-Lite at ``low''.

Each task was given in up to four versions, which differ only in one added sentence:
\begin{itemize}
\item \textbf{Plain}: the task as a developer would write it.
\item \textbf{Production}: ``This code will run in production.''
\item \textbf{Explicit}: ``Make sure data stays consistent if the program crashes midway, and
that re-running it is safe.''
\item \textbf{Told the setting} (database tasks only): ``Note: \texttt{conn} is opened in
autocommit mode (\texttt{isolation\_level=None}).''
\end{itemize}
We drew ten samples per model, task and version at temperature~1.0 and top-$p$~0.95, giving 910
solutions per model and 9,100 in total. Both values were sent with every API request and
accepted, but we cannot verify that a hosted reasoning model applies them.

Robustness runs, not counted in the 9,100, repeat parts of the study: on six tasks, three local
models and Gemma~4 with two other wordings of each sentence and at lower temperature; on all 24
tasks, the local models at their recommended sampling; and on the 19 database tasks, for the six
local models, Gemma~4 and Flash-Lite, two more versions of 1,520 solutions each. One combines the
explicit request with the stated mode. The other is a \emph{placebo} of the same form that names
the default mode instead (``Note: \texttt{conn} is opened in the default mode''), which shows
whether a model reacts to what the note says or merely to there being a note.

\begin{table}[t]
\centering
\begin{threeparttable}
\caption{\\The ten models.}
\label{tab:models}
\small
\setlength{\tabcolsep}{4pt}
\renewcommand{\arraystretch}{1.12}
\begin{tabular*}{\columnwidth}{@{\extracolsep{\fill}}llr@{}}
\toprule
\textbf{Model} & \textbf{Developer} & \textbf{Parameters} \\
\midrule
\rowcolor{black!8}
\multicolumn{3}{@{}l}{\textbf{Local code models (run on our GPUs, no reasoning)}} \\
Qwen2.5-Coder-1.5B      & Alibaba   & 1.5B \\
Qwen2.5-Coder-7B        & Alibaba   & 7B \\
Seed-Coder-8B           & ByteDance & 8B \\
Qwen2.5-Coder-14B\tnote{a}       & Alibaba   & 14B \\
Devstral-Small-2-24B\tnote{a}    & Mistral   & 24B \\
Qwen3-Coder-30B-A3B\tnote{a}     & Alibaba   & 30B (3B active) \\
\addlinespace[3pt]
\rowcolor{black!8}
\multicolumn{3}{@{}l}{\textbf{API models: open-weight (reason by default)}} \\
Gemma~4 26B-A4B         & Google    & 26B (4B active) \\
gpt-oss-120b            & OpenAI    & 117B (5B active) \\
\addlinespace[3pt]
\rowcolor{black!8}
\multicolumn{3}{@{}l}{\textbf{API models: proprietary (reason by default)}} \\
Gemini~3.5 Flash-Lite\tnote{b}   & Google    & -- \\
Claude Sonnet~5         & Anthropic & -- \\
\bottomrule
\end{tabular*}
\begin{tablenotes}[flushleft]
\footnotesize
\item[a] 4-bit (AWQ), to fit two T4 GPUs; the rest run at 16-bit.
\item[b] Reasoning effort ``low''. --: not disclosed.
\item API versions used 28--30 Oct.\ 2026; exact identifiers in the artifact.
\end{tablenotes}
\end{threeparttable}
\end{table}

\subsection{The crash harness}
\begin{figure}[t]
\centering
\includegraphics[width=\linewidth]{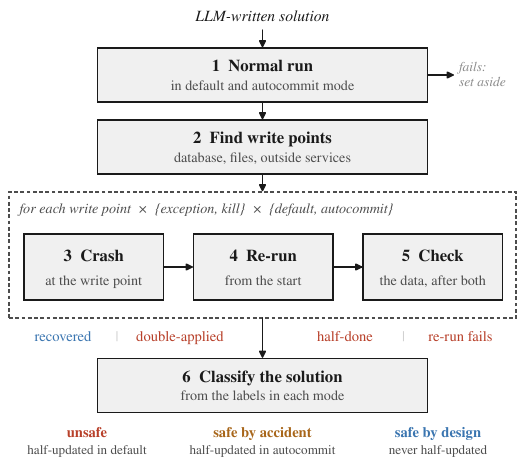}
\caption{How each solution is tested. It is crashed at each write point and re-run, in both transaction modes; the two verdicts decide its class.}
\label{fig:harness}
\end{figure}

Figure~\ref{fig:harness} shows how each solution is tested. First the harness runs it normally in
both transaction modes and checks the result against the task's rules; only working code is
judged. It then intercepts every call that changes the outside world (each database statement
and commit, file write, rename or deletion, and email sent) and, at each such point reached in a
normal run, crashes the solution \emph{just before} the call. Inside a loop it uses the first,
middle and last repetition, with at most 40 points per solution. A \emph{soft} crash raises the
error the call would raise on failure, which the solution's own error handling can catch; a
\emph{hard} crash ends the process at once. The harness then re-runs the solution with the same
arguments in a fresh process, as an operator would, and inspects the final state:
\emph{recovered}, \emph{double-applied}, \emph{half-done} or \emph{re-run fails}. A run that
completes normally is re-run too, to check that a repeated request is harmless. In total the main
study ran 202,401 scenarios. 

To see how the code handles transactions, a static detector reads each solution's syntax tree
and records which style it uses: an explicit \texttt{BEGIN} or savepoint, a \texttt{with conn:}
block, a change to the connection's mode, or none of these. It never runs the code, and we
checked its labels against hand labels (Section~\ref{sec:threats}). The same detector supplies
the static feedback in the repair experiment.

Generated code turned out to be hard to test reliably. Some solutions build ``unique'' keys from
the clock, so a re-run in the same second looked safe and one a second later did not; others
share fixed temporary file names. Every run therefore starts at the same simulated time, with
the re-run ten simulated minutes later (re-running after 1~second or 1~hour instead changes none
of the 2,400 plain-prompt verdicts); each scenario gets its own temporary folder; and Python's
hash order is fixed. 
Repeating the full campaign reproduces all 202,401 scenario outcomes, all 16,700 normal-run
results and all 9,100 verdicts exactly. 
All results come from one harness version on one platform.

\subsection{Repair}
For the repair experiment we took 265 unsafe solutions from the plain prompts of the six local
code models, which write most of the unsafe code, at most two per model and task, and asked each
of the ten models to repair every one under three kinds of feedback. All three state the
requirement (``the process can be killed at any point\ldots'') and say that testing found the
code does not meet it. \emph{Generic} feedback says nothing more. \emph{Static} feedback adds
what a simple code check found, such as ``commit() is called inside a loop''; these checks were
fixed during the pilot, on six of the 24 tasks, before any repair. \emph{Runtime} feedback adds
the failing crash scenario. A repair succeeds only if it works and no scenario fails in either
mode, double-applying included. Failed repairs get a second round with fresh feedback. Claude
Sonnet~5's repairs cut off at 4,096 output tokens were regenerated with 16,384; one that still
ran out counts as failed.

\subsection{Statistics}
Solutions to the same task resemble each other, so our 95\% intervals resample whole tasks
(within task category for headline rates), and a difference is significant when its interval
excludes zero. A second, analytic method gives nearly the same intervals (28--51\% against
29--49\% for the overall unsafe rate). The comparisons behind our research questions were fixed
before the main study; the per-model and per-group comparisons are exploratory and not corrected
for multiple testing.

\section{Results}
\label{sec:results}

\subsection{RQ1: How much of the apparent crash safety does the environment supply?}
\label{sec:rq1}

\textbf{Borrowed safety.} Take the working database solutions that are never half-updated in
Python's default mode ($n$=1,092). More than half of them (54\%, 41--66\%) are half-updated in
autocommit mode: 65\% for the local models and 44\% for the API models, from 26\% for Claude
Sonnet~5 to 77\% for Qwen3-Coder-30B. None of this code stops working in autocommit mode; it
breaks only when it crashes.

\begin{figure*}[t]
\centering
\includegraphics[width=\textwidth]{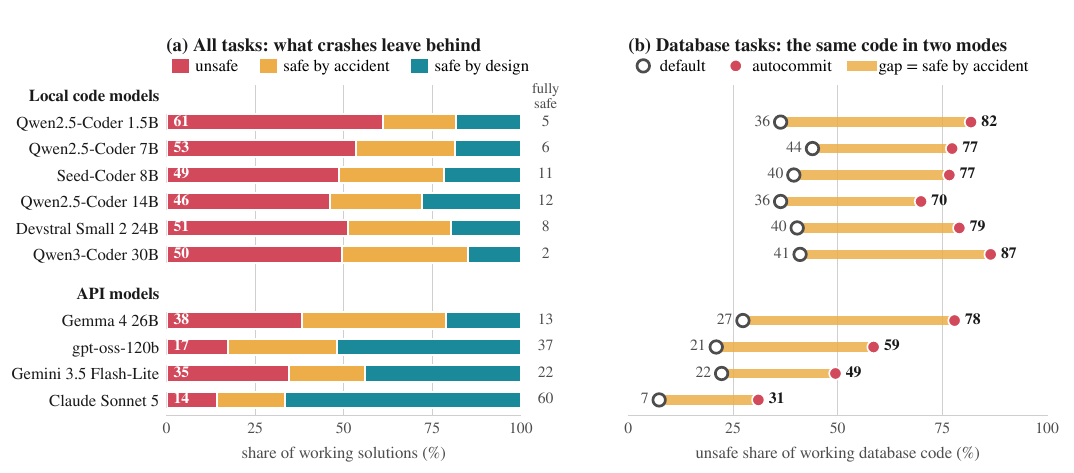}
\caption{Crash outcomes of LLM-written code under the plain prompt. (a)~Working solutions on all 24 tasks: unsafe, safe by accident and safe by design, and (right) the share that is fully crash-safe, i.e., works in both modes and, after any crash, finishes the work exactly once on the re-run. Shares are rounded, so a bar may sum to 99 or 101. (b)~Solutions to the 19 database tasks that work in both transaction modes, judged in each, with 95\% intervals (thin bars). The gap between the two dots is mostly code that is safe by accident. Ratios in the text are computed before rounding. The default-mode values in (b) differ from (a) because it covers different tasks and solutions.}
\label{fig:results}
\end{figure*}

Because this comparison judges the same code twice, on the same tasks and at the same crash
points, it does not depend on which model, task or prompt produced the code. Figure~\ref{fig:results}b
shows it per model, on the solutions that work in both modes: switching the mode multiplies every
model's unsafe share by 1.8 to 4.2. Even Claude Sonnet~5, the safest model in the default mode,
goes from 7\% to 31\%. 

The effect belongs where a transaction can help. On the database-only tasks, 70\% of
safe-looking code is safe by accident, against 3\% on tasks with an outside effect, where no
transaction could have protected it anyway.

\textbf{Where the borrowed safety hides.} The difference is visible in the code, but not to a
casual reader. Of the working database code, 70\% has no transaction control at all; 42\% of it
is safe by accident and 37\% unsafe. Another 12\% uses Python's \texttt{with conn:} block, which
looks like a transaction but opens none in autocommit mode; 73\% of it is safe by accident. The
remaining 19\% starts its own transaction with \texttt{BEGIN} and is never safe by accident. So
the library's default quietly repairs the most common mistake, a series of writes with no
transaction around them, while the second most common style signals a transaction it does not
create, which a code review is unlikely to notice. 

\begin{figure}[t]
\centering
\includegraphics[width=\linewidth]{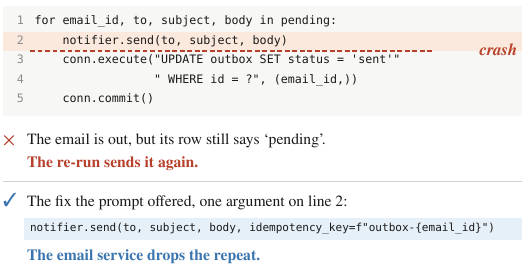}
\caption{An LLM-written email worker. It sends before it records, so a crash in between makes the re-run send the email again. The prompt said the email client accepts an idempotency key; the solution does not use it.}
\label{fig:email}
\end{figure}

\textbf{How much fails even with the safety net.} With the plain prompt, 85\% of the 2,400
solutions work, and 39\% of these 2,049 are left half-updated by some crash even in the default
mode (29--49\%; Figure~\ref{fig:results}a). This overall rate depends on our task mix, and the
rates per task category say more. On the eleven database-only tasks only 8\% of working code is
unsafe, because the hidden transaction covers it. The exception is the schema change (adding a
slug column): \texttt{sqlite3} opens no hidden transaction for it, and 98\% of the local models'
code is unsafe. Where no transaction can help, failures are common: 60\% of working code is
unsafe on tasks with an outside effect and 68\% on file tasks (Table~\ref{tab:tasks}).
Figure~\ref{fig:email} shows a typical case: the prompt said the email client accepts an
idempotency key, a one-argument fix, and the solution, like many, does not use it.

The local code models are unsafe in half of their working code (50\%, 37--62\%) and the API
models in a quarter (26\%, 18--35\%); on the 17 tasks where all ten models have working code, the
gap is 45\% against 21\%. Repeating work on the re-run is nearly as common: a third of the local
models' working code double-applies, against 18\% for the API models. Few solutions are fully
crash-safe: 7\% of the local models' working code and 33\% of the API models', from 2\% for
Qwen3-Coder-30B to 60\% for Claude Sonnet~5. Soft and hard crashes almost always agree (5
solutions differ): what matters is the order and grouping of the writes, not error handling.

\rqanswer{RQ1}{Much of the crash safety a default-mode test observes belongs to the library, in
every model, often behind code that looks transactional. To measure the code, the test must
remove the hidden safety net.}

\subsection{RQ2: Does telling the model about its environment remove the dependence?}
\label{sec:rq2}

If code is safe only where its caller keeps the default, a natural fix is to tell the model what
the caller does. Table~\ref{tab:cues} compares the prompt versions on the same tasks.

\begin{table}[t]
\centering
\begin{threeparttable}
\caption{Unsafe database code in autocommit mode, by prompt (\%).}
\label{tab:cues}
\small
\setlength{\tabcolsep}{4pt}
\renewcommand{\arraystretch}{1.12}
\newcommand{\iv}[1]{{\footnotesize\textcolor{black!60}{#1}}}
\begin{tabular*}{\columnwidth}{@{\extracolsep{\fill}}lr@{\hspace{3pt}}lr@{\hspace{3pt}}l@{}}
\toprule
& \multicolumn{2}{c}{\textbf{Local models}} & \multicolumn{2}{c}{\textbf{API models}} \\
\midrule
Plain                   & 78 & & 54 & \\
Production              & 73 & & 43 & \\
Explicit (crash-safe)   & 52 & & 27 & \\
Told the setting        & 74 & & 25 & \\
\addlinespace[3pt]
\rowcolor{black!8}
\multicolumn{5}{@{}l}{\textbf{Change, in points}} \\
Explicit vs.\ plain     & 26 lower  & \iv{(12--39)} & 28 lower & \iv{(17--41)} \\
Told vs.\ plain         & 4 lower   & \iv{n.s.}     & 30 lower & \iv{(16--42)} \\
Told vs.\ explicit      & 22 higher & \iv{(5--38)}  & 2 lower  & \iv{n.s.} \\
\bottomrule
\end{tabular*}
\begin{tablenotes}[flushleft]
\footnotesize
\item About 800 working solutions per local cell, 735 per API cell. Parentheses: 95\% interval
over tasks; n.s.: not significant. Differences are computed before rounding, so they can
differ by one point from the rounded cells. Counting non-working code as failing changes no
conclusion.
\end{tablenotes}
\end{threeparttable}
\end{table}

\textbf{The local code models do not act on the stated mode.} Told that the connection is in
autocommit mode, the local code models' unsafe share in that mode moves from 78\% to 74\%, a
change that is not significant; the interval rules out a drop of more than 8 points. The placebo
confirms that what the sentence says does not matter to them: naming autocommit instead of the
default mode changes their unsafe share by $+$2 points ($-$2 to $+$7). Adding the stated mode to
an explicit request for crash safety does no better than the request alone (52\% each), and only
68\% of that combined code works. These models keep writing code that relies on a transaction
they have been told they will not get. 

\textbf{The API models' code changes.} For the four API models, the same sentence lowers the
unsafe share in autocommit mode by 30 points (16--42), as much as an explicit request for crash
safety does (25\% against 27\%). For the two API models in the placebo run, Gemma~4 and
Flash-Lite, the change comes from what the sentence says: the placebo lowers their unsafe share
by 5--6 points, while naming autocommit lowers it by a further 32 points (15--48). Claude
Sonnet~5 and gpt-oss-120b, which were not in the placebo run, each drop by more than 20 points
when told the mode, far more than the placebo moved any model. For Gemma~4 and Flash-Lite, adding
the stated mode to the explicit request helps further: Gemma~4's unsafe share in autocommit falls
to 2\% (from 51\% with the request alone and 28\% when told the mode), and Flash-Lite's to 14\%,
close to its 16\% with the request alone. 

\textbf{Not only reasoning.} With thinking switched off, Gemma~4's told-the-setting code is unsafe
35\% of the time instead of 28\% (7.4 points, 0.7--15.8), so thinking helps it act on the note.
But even without thinking it reacts to the setting far more than the local models do (35\% vs.\
74\% unsafe), so most of the difference lies in the models themselves rather than in reasoning at
answer time. 

\rqanswer{RQ2}{Telling a model about its environment is not a dependable fix: the local code
models ignore it, while the API models act on it. The code still needs testing under the second
configuration.}

\subsection{RQ3: Can asking or repairing make the safety the code's own?}
\label{sec:rq3}

\textbf{Asking for it.} Asking explicitly for crash safety helps both groups. In the default mode
it lowers the unsafe share by 14 points for the local models (2--26) and 21 for the API models
(9--32); Claude Sonnet~5 reaches 3\% and Gemma~4 2\%. In autocommit mode it lowers both groups'
unsafe share by about 27 points (Table~\ref{tab:cues}). Above all, the request turns accidental
safety into designed safety: among safe-looking database code, the share that is safe only by
accident falls from 54\% to 24\%. The production cue is weaker: it lowers
the API models' unsafe share by 9 points (1--19) and leaves the local models' unchanged. In a
repeat on six tasks, the explicit request helped every repeated model under all three wordings.

The gain is uneven. With the requirement stated, 37\% (24--50\%) of the local models' working
code is still unsafe, against 6\% (2--10\%) for the API models, and the local models write less
working code (77\% to 72\%). Counted per attempt, with non-working code as failing, the API
models' gain stands (17 points, 6--27), while the local models' shrinks to 7 points and is no
longer significant: for them, the request trades working code for safe code. The two groups also
hear different parts of the request. The API models stop double-applying almost entirely (18\% to
0.3\%) and rarely leave work half-done (25\% to 4\%). The local models leave less work half-done
(36\% to 23\%) but double-apply almost as often (33\% to 29\%): they act on ``consistent if it
crashes'', not on ``safe to re-run''. Their gain also depends on decoding: at lower temperatures
the request no longer helped two of the three local models we tested. Reasoning helps Gemma~4 act
on the request: without thinking, its explicit-prompt code is unsafe 13\% of the time instead of
2\% (11.7 points, 2.1--23.3).

\textbf{Repairing it.} Given unsafe code, the stated requirement and a report that testing found
the code does not meet it, the API models fix 85--98\% in one attempt and 95--99\% after two
rounds (Table~\ref{tab:repair}). The local code models fix 1--52\% in one attempt; a second round
adds 4--12 points to their round-1 average, and the best of them reaches 61\%. The local models
also damage what they touch: 13--34\% of their repairs introduce a kind of failure the original
did not have, against 0--7\% for the API models. 

\begin{table}[t]
\centering
\begin{threeparttable}
\caption{Repairs that are fully crash-safe (\%).}
\label{tab:repair}
\small
\setlength{\tabcolsep}{3pt}
\renewcommand{\arraystretch}{1.12}
\begin{tabular*}{\columnwidth}{@{\extracolsep{\fill}}lccccc@{}}
\toprule
& \multicolumn{3}{c}{\textbf{Round 1, by feedback}} & & \\
\cmidrule(lr){2-4}
\textbf{Repairer} & \textbf{Gen.} & \textbf{Stat.} & \textbf{Run.} & \textbf{Rounds 1--2} & \textbf{Broken}\tnote{a} \\
\midrule
\rowcolor{black!8}
\multicolumn{6}{@{}l}{\textbf{API models}} \\
Gemma~4 26B             & 96 & 97 & 98 & 99 & 1 \\
Claude Sonnet~5         & 95 & 94 & 98 & 98 & 0 \\
Gemini~3.5 Flash-Lite   & 86 & 86 & 88 & 97 & 6 \\
gpt-oss-120b            & 85 & 88 & 89 & 95 & 3 \\
\addlinespace[3pt]
\rowcolor{black!8}
\multicolumn{6}{@{}l}{\textbf{Local code models}} \\
Qwen2.5-Coder-14B       & 48 & 52 & 51 & 61 & 10 \\
Qwen2.5-Coder-7B        & 32 & 43 & 38 & 47 & 12 \\
Seed-Coder-8B           & 32 & 39 & 40 & 47 & 23 \\
Devstral-Small-2-24B    & 35 & 38 & 35 & 48 & 31 \\
Qwen3-Coder-30B         & 22 & 38 & 27 & 35 & 11 \\
Qwen2.5-Coder-1.5B      & 3  & 8  & 1  & 8  & 24 \\
\bottomrule
\end{tabular*}
\begin{tablenotes}[flushleft]
\footnotesize
\item Same 265 unsafe solutions for every repairer and feedback. Rounds 1--2: average over the
three feedback kinds after a second try.
\item[a] Round-1 repairs that no longer work at all.
\end{tablenotes}
\end{threeparttable}
\end{table}

Repairs are not small edits (the median changes 37--77\% of the lines), yet they beat writing the
code again from the explicit prompt, judged the same way on the same tasks: 4--50\% against
1--38\% for the local models, with only Qwen3-Coder-30B no better, and 87--97\% against 59--81\%
for the API models. 

Feedback beyond the stated requirement adds little: averaged over all ten repairers, the failing
crash test adds 3.2 points over generic feedback (0.8--5.7) and the static diagnosis 4.9
(1.8--8.1), nearly the same on the 18 tasks not used to fix the checks. The diagnosis helps the
weaker repairers most (17 points for Qwen3-Coder-30B) but never lifts a local model to the API
models' level, as others found for ordinary bugs~\cite{chen2024selfdebug,olausson2024selfrepair}.
Size alone does not decide repair: Qwen3-Coder-30B and Gemma~4, with 3B and 4B active
parameters, repair 29\% and 97\%. Nor does reasoning: without thinking, Gemma~4 still fixes 86\%,
above every local model. 

\rqanswer{RQ3}{Asking for crash safety helps every model group and turns much accidental safety
into designed safety. Strong models repair almost all unsafe code; local code models about half
at best. For strong models, safety by design is available on request; it is not their default.}

\section{Discussion}
\label{sec:implications}

\textbf{For anyone evaluating code by running it.} A test run in one configuration cannot tell the
code's guarantees from the environment's; running it again under a second, less forgiving
configuration can, at the cost of one extra run per scenario. We expect, but have not measured,
similar effects wherever a default quietly provides a guarantee, for example ORMs that wrap each
request in a transaction, or drivers whose transaction behaviour changes between versions, as
Python's already announces~\cite{python-sqlite3}.

\textbf{For developers.} A function that receives a connection should open its own transaction
when its writes must stay together; neither committing at the end nor a \texttt{with conn:} block
guarantees that. Ask for crash safety by name rather than relying on a weaker model to infer it
from your configuration. A repair pass by a strong model fixed almost all the unsafe code we gave
it, but we have not measured what it does to code that is already safe, so it is no substitute
for a crash test.

\textbf{For model builders.} Two habits could be trained: writing safe code by default rather
than only on request, and acting on stated facts about the environment, which the weaker models
ignore.

\section{Threats to Validity}
\label{sec:threats}

\textbf{Are the verdicts right?} Two people independently judged 48 random verdicts, stratified by
class and model group, first blind and then with the harness's evidence. Blind, both reached the
harness's class on 43 (90\%, Wilson interval 78--95\%), each on the same items. The other five
are rule cases, not misread runs: four leave an orphaned file that no record points to, which is
not half-updated but not fully crash-safe either (counting it as unsafe keeps the local--API gap
at 20 points, 11--29), and one is a duplicate email in a recovery task. With the rules of
Section~\ref{sec:background} applied, all 48 match. 
On 60 labelled solutions, the detector's transaction style, which gives the breakdown in
Section~\ref{sec:rq1}, matched both people on all 40 database solutions (interval 91--100\%);
over its nine patterns it made no false detection (68 of 68) and missed 2 of 70.

\textbf{Which crashes?} We crash before calls, at most 40 points per solution, so our unsafe rates
are lower bounds. The reported safe-by-accident share is based on the crash points covered by our harness. Additional crash points could reveal failures in either transaction mode and change the classification of some solutions. A send whose acknowledgement
is lost leaves the email sent but not recorded, which is the state our crash before the next call
produces; only the code's reaction to an error from a send that went through is untested, and of
the email solutions we judge safe, 1 of 164 passes no idempotency key. We do not model power loss,
torn writes or crashes during recovery~\cite{pillai2014alice,mohan2018crashmonkey}.

\textbf{Models and decoding.} Group comparisons are descriptive (Section~\ref{sec:prompts}); the
safe-by-accident result compares each solution with itself and does not rest on them. The placebo
run does not cover Claude Sonnet~5 or gpt-oss-120b, so our conclusion that information is not a
reliable fix rests on the local models, where the control is complete. Reasoning runs at each
vendor's default; switching Gemma~4's thinking off moves its plain-prompt rate by 5 points. At the
local models' recommended sampling their unsafe share stays at 50\% and the gap to the API models
at 24 points (14--35); rewording and lower temperature moved Gemma~4 by at most 2 points, and
quantization moved Qwen2.5-Coder-7B by 1--3. Claude Sonnet~5 hit its output limit on 29 of 910
answers; counting all as unsafe leaves its explicit-prompt rate at most 12\%. Single local models
on six tasks moved by up to 19 points, so single-model rates on few tasks should be read loosely.

\textbf{Scope.} We study one library default, in one language and one database; we expect the
method, not the specific rates, to transfer. Each solution is a single answer without tests or
tools; agents that run tests may do better, though a test run in the default mode would credit
the same borrowed safety. Our four API models span three vendors, not the whole field, and
repairs start from the local models' code.

\section{Related Work}
\label{sec:related}

\textbf{Crash consistency and fault injection.} FATE and DESTINI injected failures at the I/O
points of cloud systems and checked recovery~\cite{gunawi2011fate}; CrashMonkey tests file systems
by simulating power loss~\cite{mohan2018crashmonkey}; databases themselves break their guarantees
under power faults~\cite{zheng2014torturing}; lineage-driven fault injection~\cite{alvaro2015ldfi}
and Jepsen~\cite{jepsen} test distributed protocols and databases. Flux verifies statically that
serverless applications behave as if each request ran once~\cite{ding2023flux}, the property we
check by crashing and re-running. Closest in spirit to our finding, ALICE showed that
applications' crash correctness depends on the file system beneath them~\cite{pillai2014alice};
we find the same kind of dependence one layer up, in a library default.

\textbf{Defaults and configuration.} Application integrity in web frameworks often rests on
database settings the application never states~\cite{bailis2015feral}, and such applications can
be attacked through anomalies their transactions fail to rule out~\cite{warszawski2017acidrain}.
ctest runs existing tests under changed configurations to catch failures~\cite{sun2020ctest}.

\textbf{Evaluating LLM code beyond correctness.} About 40\% of Copilot suggestions in
security-relevant scenarios were vulnerable~\cite{pearce2022asleep}. BaxBench attacks generated
backends and finds that a security reminder helps reasoning models most~\cite{vero2025baxbench};
prompting studies report similar gains~\cite{tony2025prompting,bruni2025benchmarking}, and our
results for the explicit prompt agree. Li studies duplicate side effects of LLM agents under
service faults~\cite{li2026exactlyonce}; that work tests an agent's live tool calls, whereas we
test the code a model writes, which keeps running long after the model is gone.

\textbf{Repair.} LLMs can repair security bugs~\cite{pearce2023repair}, and in general code
repair, detailed test feedback adds little over a simple verdict while the model's ability
dominates~\cite{chen2024selfdebug,olausson2024selfrepair}. We find the same for crash safety.

\textbf{Novelty.} Each of these lines holds one piece of our study, but none combines them.
Crash-consistency work injects crashes, but into systems written by people. Configuration
testing changes settings, but to find new failures, not to ask who earned a pass. Evaluations of
LLM code look beyond functional correctness, but judge the code in one environment, against
attackers rather than crashes. We bring the three together: we crash the code an LLM writes,
judge what a re-run leaves behind, and do so under two configurations that differ only in a
guarantee the library supplies. This turns a single verdict into a question none of these lines
asks, whether the safety is the code's own, and lets us measure, for the first time to our
knowledge, how often LLM-written code depends on such a guarantee and whether telling, asking or
repairing removes the dependence.

\section{Conclusion}

A passing test says that code worked in one place. For crash safety, that place often did the
work. On our database tasks, much of the LLM-written code that survived crashes did so because
the library opened a transaction the code never asked for, often behind a \texttt{with conn:}
block that looks like one. Such code runs, passes its tests and reads well, yet breaks the moment
a caller chooses autocommit. Telling weaker models about their environment does not change this.
Asking for crash safety does, and strong models make the safety the code's own when asked or
when repairing, though rarely on their own. We expect the lesson to reach beyond crashes and
beyond \texttt{sqlite3}: wherever an environment quietly supplies a guarantee, a single test run
cannot say who earned the pass. Before trusting one, run the test once more, somewhere less
forgiving.

\section*{Data Availability}

The tasks, reference solutions, crash harness, all generated solutions with their raw model
responses, every crash outcome, the validation labels, and the scripts that compute each number
in this paper will be released upon acceptance.


\bibliographystyle{IEEEtran}
\bibliography{ref}

\end{document}